\documentclass[aps,english,showpacs,twocolumn]{revtex4-1}
\usepackage{amssymb}
\usepackage{amsmath}
\usepackage{graphicx}
\usepackage{epsfig}

\begin{document}

\title{Hermitian dynamics in a class of pseudo-Hermitian networks}
\author{L. Jin and Z. Song}
\email{songtc@nankai.edu.cn}
\affiliation{School of Physics, Nankai University, Tianjin 300071, China}

\begin{abstract}
{We investigate a pseudo-Hermitian lattice system, which consists of a set
of isomorphic pseudo-Hermitian clusters coupled together in a Hermitian
manner. We show that such non-Hermitian systems can act as Hermitian
systems. This is made possible by considering the dynamics of a state
involving an identical eigenmode of each isomorphic cluster. It still holds
when multiple eigenmodes are involved when additional restriction on the
state is imposed. This Hermitian dynamics is demonstrated for the case of an
exactly solvable $\mathcal{PT}$-symmetric ladder system.}
\end{abstract}

\maketitle


\section{Introduction}

The Hermitian quantum mechanics is a well-developed framework because a
Hermitian Hamiltonian leads to a real spectrum and unitary time evolution,
which preserves the probability normalization. However, a decade ago it was
observed that a large class of non-Hermitian Hamiltonians possess real
spectra \cite{Bender98,IX} and\ a pseudo-Hermitian Hamiltonian connects with
its equivalent Hermitian Hamiltonian via a similarity transformation \cite%
{AM43,AM37}, quantum theory based on non-Hermitian Hamiltonian was
established \cite{AM37,Ahmed,Berry,Heiss,Jones,Muga,Bender02,BenderRPP,Tateo,AMIJGMMP}. In
additional, such a framework also indicates the preservation of probability
normalization if a positive-definite inner product is employed to replace
the Dirac inner product.\ Nevertheless, to date the interpretation and the
measurement in experiment of such an inner product are not clear. While the
Dirac probability (Dirac inner product) can be measured in a universal
manner. Therefore being an acceptable theory of quantum mechanics, the Dirac
probability is of central importance to most practical physical problems.

Parity and time-reversal symmetric ($\mathcal{PT}$-symmetric) system has attracted much attention due to recent
progresses on experimental investigations in $\mathcal{PT}$-symmetric
optical systems, observation of passive $\mathcal{PT}$-symmetric breaking in
passive optical double-well structure \cite{AGuo} and observation of
spontaneous symmetry breaking together with power oscillations in optical
coupled system \cite{Ruter} were carried out. Optics offers the rather
unique advantage in detection the wave function evolution and seem to be the
most readily applicable \cite{LonghiRev,Klaiman}. In the past two decades,
general issues of quantum effects in quantum systems have proven to be
successfully investigated in the framework of quantum optical analogy based
on the fact that paraxial propagation of light in optical guided structures
is governed by a Schr\"{o}dinger-like equation \cite{LonghiRev}. Actually,
the intensity observed in optical experiment corresponds to the Dirac
probability of electric field envelope. It is not guaranteed for generic
systems that the Dirac probability is preserved even when the non-Hermitian
Hamiltonian is time independent. Nevertheless, the violation of the
conservation of Dirac probability\ in non-Hermitian system does not
contradict the Copenhagen interpretation. The implications of
pseudo-Hermitian system are still under consideration,\ peculiar features\
were exhibited such as double refraction, power oscillations, \textit{etc.}
\cite{PRL103904,PRL030402} following by the suggestion of realization of $%
\mathcal{PT}$-symmetric structure in the realm of optics \cite{OL2632},
while nonreciprocal Bloch oscillation with no classical correspondence was
also shown in $\mathcal{PT}$-symmetric complex crystal \cite{LonghiPRL}.

We propose a class of non-Hermitian lattice systems in this work, the system
is composed by a set of isomorphic pseudo-Hermitian clusters, which
connected with each other in a Hermitian way. We show that in such
non-Hermitian systems, Hermitian like\ dynamics can be observed, including
the property that the time evolution is Dirac probability preserving. This
is made possible by considering the dynamics of a state involving the
superposition of an identical eigenmode of each isomorphic cluster in
general case. In the case of having additional orthogonal modes, it still
holds when multiple eigenmodes are involved. This Hermitian dynamics, as
well as the quasi-Hermitian behavior, are specifically demonstrated for the
case of an exactly solvable pseudo-Hermitian system.

This paper is organized as follows. In Section \ref{sec_H_Basic}, we present
the model we focus on and its basic properties. Section \ref%
{sec_Pseudo_H_ladder} consists of an exactly solvable example to illustrate
our main idea. Section \ref{sec_conclusion} is the summary and discussion.

\section{Hamiltonian and basic properties}

\label{sec_H_Basic} A general tight-binding network is constructed
topologically by the sites and the various connections between them. As a
simplified model, it captures the essential features of many discrete
systems. Also it is a nice testing ground for the study of the non-Hermitian
quantum mechanics due to its analytical and numerical tractability. Much
effort has been devoted in recent years to the pseudo-Hermitian lattice system \cite%
{MZnojil,Bendix,Longhi,Joglekar,Joglekar83,Weston,Fring,Ghosh,Giorgi,ZLi,Bousmina,Ozlem,Fabio}%
. The Hamiltonian of the concerned tight-binding network reads as follows

\begin{eqnarray}
H &=&\sum_{\alpha =1}^{N}H_{\alpha }+\sum_{\alpha <\beta }H_{\alpha \beta },
\label{H} \\
H_{\alpha } &=&\lambda _{\alpha }\sum_{i,j=1}^{N_{d}}J_{ij}a_{\alpha
,i}^{\dag }a_{\alpha ,j}, \\
H_{\alpha \beta } &=&\kappa _{\alpha \beta }\sum_{l=1}^{N_{d}}a_{\alpha
,l}^{\dag }a_{\beta ,l}+\text{H.c.},
\end{eqnarray}%
which consists of $N$ isomorphic clusters $H_{\alpha }$, with each cluster has a dimension
$N_d$. The label $\alpha $ denotes the $\alpha $th subgraph of $N$ clusters, and
 $a_{\alpha ,i}^{\dag }$\ ($a_{\alpha ,i}$) is the boson or fermion creation (annihilation) operator at
the $i$th site in the $\alpha $th cluster. The cluster $H_{\alpha }$ is
defined by the distribution of the hopping integrals $\left\{ \lambda
_{\alpha }J_{ij}\right\} $ where $\lambda _{\alpha }$\ is real. The set of
clusters are isomorphic due the fact that they have the same eigenfunctions
and spectral structures. Note that terms $\sum_{\alpha <\beta }H_{\alpha
\beta }$ is self-adjoint since $H_{\alpha \beta }=H_{\alpha \beta }^{\dag }$%
, which describes the Hermitian connection between clusters.
And such kind of couplings are the type of \textit{similarity mapping}, which
is crucial for the conclusion of this paper. The total Hamiltonian $H$\ is
not Hermitian when the matrix $J_{ij}$ is not Hermitian. Figure \ref%
{fig_illus}(a) shows a schematic example.


\begin{figure}[tbp]
\includegraphics[ bb=57 460 539 733, width=8.5 cm, clip]{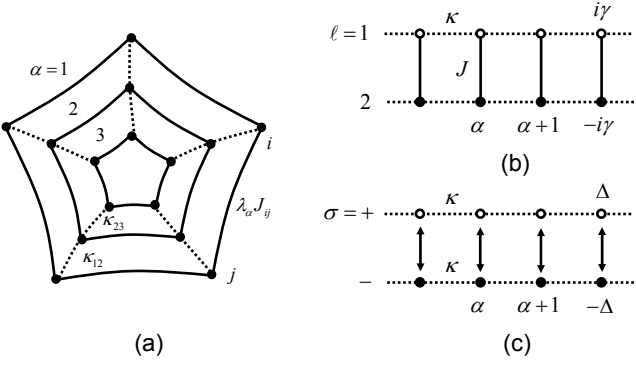}
\caption{Schematic illustration of the concerned networks. (a) A lattice
consists of three $5$-site isomorphic clusters, where the different sizes
indicate the factor $\lambda _{\alpha }$.\ The dot lines
denote the similarity-mapping-type Hermitian structure couplings across the clusters. (b) A
concrete example which is a two-leg ladder. Each rung is a non-Hermitian
cluster. (c) Equivalent two-band model Eq. (\ref{H_Ladd_2band}).
Here the double-headed arrow denotes the quasi-canonical commutation relations
between the eigenmodes $\sigma =\pm $\ for the same cluster.} \label%
{fig_illus}
\end{figure}


In this paper, we consider the case of $H_{\alpha }$ being pseudo-Hermitian,
i.e., $H_{\alpha }$ is non-Hermitian but has entirely real spectrum. Then $H$%
\ is also pseudo-Hermitian in the case of real $\lambda _{\alpha }$, hence
possessing the common exceptional point as $H_{\alpha }$.\ In general, a
pseudo-Hermitian Hamiltonian does not guarantee the Dirac probability
preserving. It has been shown that the Dirac norm of an evolved wavepacket
ceases preserving as long as it touches the region of on-site imaginary potentials
\cite{LJinCTP}. In the following we will show\ due to the pseudo-Hermitian
clusters combined together in a Hermitian way, that there exist quantum
states obeying Dirac probability preserving, even if their profiles cover
the imaginary potentials.

We start with the eigen problem of the Hamiltonian $H_{\alpha }$. In
single-particle invariant subspace, following the well established
pseudo-Hermitian quantum mechanics \cite{BenderRPP,Tateo,AMIJGMMP}, we
always have

\begin{equation}
H_{\alpha }\bar{a}_{\alpha ,\sigma }\left\vert \text{vac}\right\rangle
=\lambda _{\alpha }\epsilon _{\sigma }\bar{a}_{\alpha ,\sigma }\left\vert
\text{vac}\right\rangle ,  \label{Schrodinger_Eq_H}
\end{equation}%
and

\begin{equation}
H_{\alpha }^{\dagger }a_{\alpha ,\sigma }^{\dag }\left\vert \text{vac}%
\right\rangle =\lambda _{\alpha }\epsilon _{\sigma }a_{\alpha ,\sigma
}^{\dag }\left\vert \text{vac}\right\rangle ,  \label{S_Eq_H_dag}
\end{equation}%
where $\alpha \in \left[ 1,N\right] $\ and $\sigma \in \left[ 1,N_{d}\right]
$, the operators $\bar{a}_{\alpha ,\sigma }$ and $a_{\alpha ,\sigma }$ have
the form%
\begin{equation}
\bar{a}_{\alpha ,\sigma }=\sum_{l}f_{l\sigma }a_{\alpha ,l}^{\dag },\text{ }%
a_{\alpha ,\sigma }=\sum_{l}g_{l\sigma }^{\ast }a_{\alpha ,l},
\label{can pair}
\end{equation}%
where
\begin{equation}
\sum_{\sigma }g_{l\sigma }^{\ast }f_{l^{\prime }\sigma }=\delta _{ll^{\prime
}},\sum_{l}g_{l\sigma }^{\ast }f_{l\sigma ^{\prime }}=\delta _{\sigma \sigma
^{\prime }}.  \label{orth-re}
\end{equation}%
Note that $\left\{ f_{l\sigma }\right\} $, $\left\{ g_{l\sigma }\right\} $\
and $\left\{ \varepsilon _{\sigma }\right\} $\ are independent of $\alpha $.
Then the operators $\bar{a}_{\alpha ,\sigma }$ and $a_{\alpha ,\sigma }$ are
canonical conjugate pairs, satisfying%
\begin{eqnarray}
\lbrack a_{\alpha ,\sigma },\bar{a}_{\alpha ^{\prime },\sigma ^{\prime
}}]_{\pm } &=&\delta _{\alpha \alpha ^{\prime }}\delta _{\sigma \sigma
^{\prime }},  \label{can CR} \\
\lbrack a_{\alpha ,\sigma },a_{\alpha ^{\prime },\sigma ^{\prime }}]_{\pm }
&=&[\bar{a}_{\alpha ,\sigma },\bar{a}_{\alpha ^{\prime },\sigma ^{\prime
}}]_{\pm }=0,
\end{eqnarray}%
where $[\cdot ,\cdot ]_{\pm }$ denotes the the commutator and
anti-commutator. And accordingly, the original Hamiltonian can be rewritten
as the form

\begin{eqnarray}
H &=&\sum_{\alpha ,\sigma }\lambda _{\alpha }\epsilon _{\sigma }\bar{a}%
_{\alpha ,\sigma }a_{\alpha ,\sigma }  \label{H_Sigma} \\
&&+\sum_{\alpha <\beta ,\sigma }\left( \kappa _{\alpha \beta }\bar{a}%
_{\alpha ,\sigma }a_{\beta ,\sigma }+\kappa _{\alpha \beta }^{\ast }\bar{a}%
_{\beta ,\sigma }a_{\alpha ,\sigma }\right) ,  \notag
\end{eqnarray}%
which has the following subtle features: (i) The matrix representation of $H$%
\ with respect to the biorthogonal basis \{$\left\langle \text{vac}%
\right\vert a_{\alpha ,\sigma },\bar{a}_{\alpha ,\sigma }\left\vert \text{vac%
}\right\rangle $\} is Hermitian, i.e., $\left\langle \text{vac}\right\vert
a_{\alpha ,\sigma }H\bar{a}_{\alpha ^{\prime },\sigma ^{\prime }}\left\vert
\text{vac}\right\rangle =$ $\left( \left\langle \text{vac}\right\vert
a_{\alpha ^{\prime },\sigma ^{\prime }}H\bar{a}_{\alpha ,\sigma }\left\vert
\text{vac}\right\rangle \right) ^{\ast }$; (ii) Although it is a
non-Hermitian operator, i.e., $H\neq H^{\dagger }$, straightforward algebra
shows that

\begin{equation}
\lbrack a_{\alpha ,\sigma },a_{\alpha ^{\prime },\sigma ^{\prime }}^{\dag
}]_{\pm }\propto \delta _{\alpha \alpha ^{\prime }},\text{ }[a_{\alpha
,\sigma },a_{\alpha ^{\prime },\sigma ^{\prime }}]_{\pm }=0,
\label{quasi CR}
\end{equation}%
which indicates that although with nonorthogonality of the eigenstates as
inherent feature of non-Hermitian system, $a_{\alpha ,\sigma }$ and $%
a_{\alpha ^{\prime },\sigma ^{\prime }}^{\dag }$ obey quasi-canonical
commutation relations due to the Hermitian connection structure between
clusters. This results in a new type of particle statistics, that is rarely
observed in Hermitian systems, thus becomes highly relevant in the presence
of non-Hermitian terms.

Considering an arbitrary state in the form%
\begin{equation}
\left\vert \Phi _{\sigma }\left( 0\right) \right\rangle =\sum_{\alpha
}c_{\alpha }\bar{a}_{\alpha ,\sigma }\left\vert \text{vac}\right\rangle ,
\label{arb ini}
\end{equation}%
as the initial state, where $\sum_{\alpha }\left\vert c_{\alpha }\right\vert
^{2}=1$ and in which only the eigenmode $\sigma $ of each cluster is
involved. At instant $t$, we have
\begin{equation}
\left\vert \Phi _{\sigma }\left( t\right) \right\rangle =\sum_{\alpha
}c_{\alpha }e^{-iHt}\bar{a}_{\alpha ,\sigma }\left\vert \text{vac}%
\right\rangle .  \label{phi_t}
\end{equation}%
In the framework of metric operator theory, $H$\ acts as a Hermitian system,
obeying unitary time evolution in the positive-definite inner product \cite%
{AM37}. However to date the physical meaning of the positive-definite inner
product is unclear, while the Dirac probability\ can be measured in a
universal manner, e.g. Dirac probability of wave electric field corresponds
to the light intensity in optics and is simple to detect in experiment \cite%
{LonghiRev}, therefore Dirac norm is of central importance. The aim of this
paper is to show that contrary to\ the nonclassical dynamical behavior \cite%
{PRL103904,LonghiPRL}, the unitary Dirac probability dynamics can also be
observed in the pseudo-Hermitian system. Actually inserting $\sum_{\beta
,\sigma ^{\prime }}\bar{a}_{\beta ,\sigma ^{\prime }}\left\vert \text{vac}%
\right\rangle \left\langle \text{vac}\right\vert a_{\beta ,\sigma ^{\prime
}}=1$\ into Eq. (\ref{phi_t}), we have

\begin{eqnarray}
\left\vert \Phi _{\sigma }\left( t\right) \right\rangle &=&\sum_{\alpha
,\beta }c_{\alpha }\bar{a}_{\beta ,\sigma }\left\vert \text{vac}%
\right\rangle \left\langle \text{vac}\right\vert a_{\beta ,\sigma }e^{-iHt}%
\bar{a}_{\alpha ,\sigma }\left\vert \text{vac}\right\rangle  \notag \\
&=&\sum_{\alpha ,\beta }c_{\alpha }U_{\beta \alpha }\bar{a}_{\beta ,\sigma
}\left\vert \text{vac}\right\rangle ,  \label{phi_t_U}
\end{eqnarray}%
where%
\begin{equation}
U_{\beta \alpha }=\left\langle \text{vac}\right\vert a_{\beta ,\sigma
}e^{-iHt}\bar{a}_{\alpha ,\sigma }\left\vert \text{vac}\right\rangle ,
\label{U}
\end{equation}%
is the propagator in the framework of biorthogonal\ basis and satisfies%
\begin{equation}
\sum_{\gamma }U_{\gamma \alpha }U_{\gamma \beta }^{\ast }=\delta _{\alpha
\beta },  \label{Unitary}
\end{equation}%
due to the above mentioned feature (i) of $H$. Accordingly, the Dirac norm
has the form%
\begin{eqnarray}
&&\left\vert \left\vert \Phi _{\sigma }\left( t\right) \right\rangle
\right\vert ^{2}=\left( \left\vert \Phi _{\sigma }\left( t\right)
\right\rangle \right) ^{\dag }\left\vert \Phi _{\sigma }\left( t\right)
\right\rangle  \label{Dirac norm1} \\
&=&(\sum_{\alpha ^{\prime },\beta ^{\prime }}c_{\alpha ^{\prime }}^{\ast
}U_{\beta ^{\prime }\alpha ^{\prime }}^{\ast }\left\langle \text{vac}%
\right\vert \bar{a}_{\beta ^{\prime },\sigma }^{\dagger })(\sum_{\alpha
,\beta }c_{\alpha }U_{\beta \alpha }\bar{a}_{\beta ,\sigma }\left\vert \text{%
vac}\right\rangle )  \notag \\
&=&\sum_{\alpha }\left\vert c_{\alpha }\right\vert ^{2}\Delta _{\sigma
}=\Delta _{\sigma },  \notag
\end{eqnarray}%
where the relation Eq. (\ref{Unitary}) is applied and the\ $\alpha $%
-independent factor $\Delta _{\sigma }$ can be obtained from%
\begin{equation}
\left\langle \text{vac}\right\vert \bar{a}_{\alpha ,\sigma }^{\dagger }\bar{a%
}_{\beta ,\sigma }\left\vert \text{vac}\right\rangle =\Delta _{\sigma
}\delta _{\alpha \beta }.  \label{Delta}
\end{equation}%
It follows that although $\left\vert \Phi _{\sigma }\left( t\right)
\right\rangle $ is not the eigenstate of the entire network system, the time
evolution is Dirac norm-conserving, this is a direct consequence of the
quasi-canonical commutation relations. The result presented here for the
evolution of an arbitrary state involving an identical
isomorphic-cluster-eigenmode provides a new way for connecting the
pseudo-Hermitian and Hermitian systems.

It is worth to mention that this probability preserving evolution can also
occur for a state involving multiple eigenmodes. This due to the fact that
there always exist states,\ which parts belong to different eigenmodes are
orthogonal in terms of Dirac inner product, hence preserve the Dirac
probability. For instance, a state involves two eigenmodes $\sigma _{1}$ and
$\sigma _{2}$, its parts on $\sigma _{1}$\ and $\sigma _{2}$ are\ spatially
separated local states with respect to the coordinate space $\alpha $, then
the two parts of the state are orthogonal in terms of Dirac inner product
and the evolution of such a state is probability preserving since the
quasi-canonical commutation relations. We will demonstrate this point
explicitly via the following illustrative example.

\section{Pseudo-Hermitian ladder}

\label{sec_Pseudo_H_ladder} Now we investigate a concrete example to
demonstrate the application of the previous result. We consider a system of
a\ two-leg ladder [Fig. \ref{fig_illus}(b)], consisting of $N$ dimers as
pseudo-Hermitian clusters. The Hamiltonian reads%
\begin{eqnarray}
&&H_{\text{Ladd}}=\sum_{\alpha =1}^{N}H_{\alpha }+\sum_{\alpha
=1}^{N}H_{\alpha ,\alpha +1},  \label{H_ladd} \\
&&H_{\alpha }=-J(a_{\alpha ,1}^{\dag }a_{\alpha ,2}+\text{H.c.})+i\gamma
\left( n_{\alpha ,1}-n_{\alpha ,2}\right) , \\
&&H_{\alpha ,\alpha +1}=-\kappa \sum_{\ell =1}^{2}(a_{\alpha ,\ell }^{\dag
}a_{\alpha +1,\ell }+\text{H.c.}),
\end{eqnarray}%
where $n_{\alpha ,\ell }=a_{\alpha ,\ell }^{\dag }a_{\alpha ,\ell }$ is the
particle number operator and the operators obey the periodic boundary
condition $a_{N+1,\ell }^{\dag }=a_{1,\ell }^{\dag },$ with $\ell =1$, $2$. $\kappa $
($J$) is the hopping integral along legs (rungs) and\ $\gamma $
denotes the norm of the imaginary on-site potential. Note that the ladder is
a $\mathcal{PT}$-symmetric Hamiltonian, where $\mathcal{P}$ is the parity
and $\mathcal{T}$ denotes time-reversal. The simple structure of this model
makes it an ideal testing ground for a more profound understanding of the
Hermitian dynamics in a pseudo-Hermitian system.\ Taking the transformations%
\begin{eqnarray}
\bar{a}_{\alpha ,\sigma } &=&\frac{1}{\sqrt{2\cos \theta }}\left( e^{i\sigma
\theta /2}a_{\alpha ,1}^{\dagger }-\sigma e^{-i\sigma \theta /2}a_{\alpha
,2}^{\dagger }\right) ,  \label{Can_ope} \\
a_{\alpha ,\sigma } &=&\frac{1}{\sqrt{2\cos \theta }}\left( e^{i\sigma
\theta /2}a_{\alpha ,1}-\sigma e^{-i\sigma \theta /2}a_{\alpha ,2}\right) ,
\end{eqnarray}
where $\left( \alpha \in \left[ 1,N\right] \text{, }\sigma =\pm \right) $,
which are obtained from the solution of the dimer (a general solution of $%
N_{d}$-dimension cluster is shown in Ref. \cite{LJin80}.), the ladder
Hamiltonian can be written as
\begin{eqnarray}
H_{\text{Ladd}} &=&\sum_{\alpha =1,\sigma =\pm }^{N}(-\kappa \bar{a}_{\alpha
,\sigma }a_{\alpha +1,\sigma }-\kappa \bar{a}_{\alpha +1,\sigma }a_{\alpha
,\sigma }  \label{H_Ladd_2band} \\
&&+\sigma \Delta \bar{a}_{\alpha ,\sigma }a_{\alpha ,\sigma }).  \notag
\end{eqnarray}%
which is illustrated in Fig. \ref{fig_illus}(c), here $\Delta =\sqrt{%
J^{2}-\gamma ^{2}}$ and $\sin \theta =\gamma /J$, $\theta \in \left[ 0,\pi /2%
\right] $. The biorthogonal structure of the solution for a dimer admits the
following canonical commutation relations Eq. (\ref{can CR}) and
\begin{eqnarray}
\left[ \bar{a}_{\alpha ,\sigma }^{\dagger },\bar{a}_{\alpha ^{\prime
},\sigma }\right] _{\pm } &=&\left[ a_{\alpha ,\sigma },a_{\alpha ,\sigma
}^{\dagger }\right] _{\pm }=\sec \theta \delta _{\alpha \alpha ^{\prime }},
\\
\lbrack \bar{a}_{\alpha ,-\sigma }^{\dagger },\bar{a}_{\alpha ^{\prime
},\sigma }] _{\pm } &=&\lbrack a_{\alpha ,\sigma },a_{\alpha ,-\sigma
}^{\dagger }] _{\pm }=i\sigma \tan \theta \delta _{\alpha \alpha
^{\prime }}.
\end{eqnarray}%
Obviously,
Hamiltonian Eq. (\ref{H_Ladd_2band}) represents a two-band model, which has
an interesting feature comparing to a Hermitian two-band model: although
there are no interband transitions, the two bands are not independent. It is
due to the pseudo-Hermiticity of the clusters, which allows $\left[
a_{\alpha ,\sigma },\bar{a}_{\alpha ,-\sigma }\right] _{\pm }=0$\ but $%
[a_{\alpha ,\sigma },a_{\alpha ,-\sigma }^{\dag }]_{\pm }\neq 0$. This
characteristic will be further demonstrated through the following
quasi-canonical commutation relations Eq. (\ref{commu_rela}) and the time
evolution for various Gaussian wavepackets.\ Figure \ref{fig_illus}(c)
schematically illustrates such an equivalent two-band structure.
Nevertheless, Hamiltonian Eq. (\ref{H_Ladd_2band}) can be diagonalized as a
Hermitian one, i.e., we have%
\begin{eqnarray}
H_{\text{Ladd}} &=&\sum_{k,\sigma =\pm }\varepsilon _{k,\sigma }\bar{a}%
_{k,\sigma }a_{k,\sigma }, \\
\varepsilon _{k,\pm } &=&-2\kappa \cos k\pm \Delta ,
\end{eqnarray}%
by using\ the linear transformations%
\begin{eqnarray}
\bar{a}_{k,\sigma } &=&\frac{1}{\sqrt{N}}\sum_{j=1}^{N}e^{ikj}\bar{a}%
_{j,\sigma }, \\
a_{k,\sigma } &=&\frac{1}{\sqrt{N}}\sum_{j=1}^{N}e^{-ikj}a_{j,\sigma },
\end{eqnarray}%
where $k=2n\pi /N$, $n\in \left[ 1,N\right] $. The linearity of the
transformations allows

\begin{eqnarray}
\left[ a_{k,\sigma },\bar{a}_{k^{\prime },\sigma ^{\prime }}\right] _{\pm }
&=&\delta _{kk^{\prime }}\delta _{\sigma \sigma ^{\prime }},
\label{can CR k} \\
\left[ \bar{a}_{k,\sigma },\bar{a}_{k^{\prime },\sigma ^{\prime }}\right]
_{\pm } &=&\left[ a_{k,\sigma },a_{k^{\prime },\sigma ^{\prime }}\right]
_{\pm }=0.
\end{eqnarray}%
However, when dealing with the Dirac inner product, the quasi-canonical commutation relations
\begin{subequations}
\label{commu_rela}
\begin{eqnarray}
\lbrack \bar{a}_{k,\sigma }^{\dagger },\bar{a}_{k^{\prime },\sigma }]_{\pm }
&=&[a_{k,\sigma },a_{k^{\prime },\sigma }^{\dagger }]_{\pm }=\sec \theta
\delta _{kk^{\prime }},  \label{commu_rela_a} \\
\lbrack \bar{a}_{k,-\sigma }^{\dagger },\bar{a}_{k^{\prime },\sigma }]_{\pm
} &=&[a_{k,\sigma },a_{k^{\prime },-\sigma }^{\dagger }]_{\pm }=i\sigma \tan
\theta \delta _{kk^{\prime }},  \label{commu_rela_b}
\end{eqnarray}
\end{subequations}
will be taken into account. Such quasi-canonical commutation relations
reflect the subtle features of the system: when dealing with different $k$, $%
a_{k,\sigma }$ and $a_{k^{\prime },\sigma ^{\prime }}^{\dagger }$ act as
canonical conjugate pairs and the system displays Hermitian behavior.

We can gain some insight regarding the role of the quasi-canonical
statistics. We will see shortly that such a model displays the similar
dynamics as a Hermitian ladder. We start our investigation from the quantum
dynamics of various initial wavepackets. In the situation of a Hermitian
ladder, any two wavepackets propagate independently and the total
probability is preserving.


\begin{figure*}[tbp]
\includegraphics[ bb=30 185 403 601, width=6.5 cm, clip]{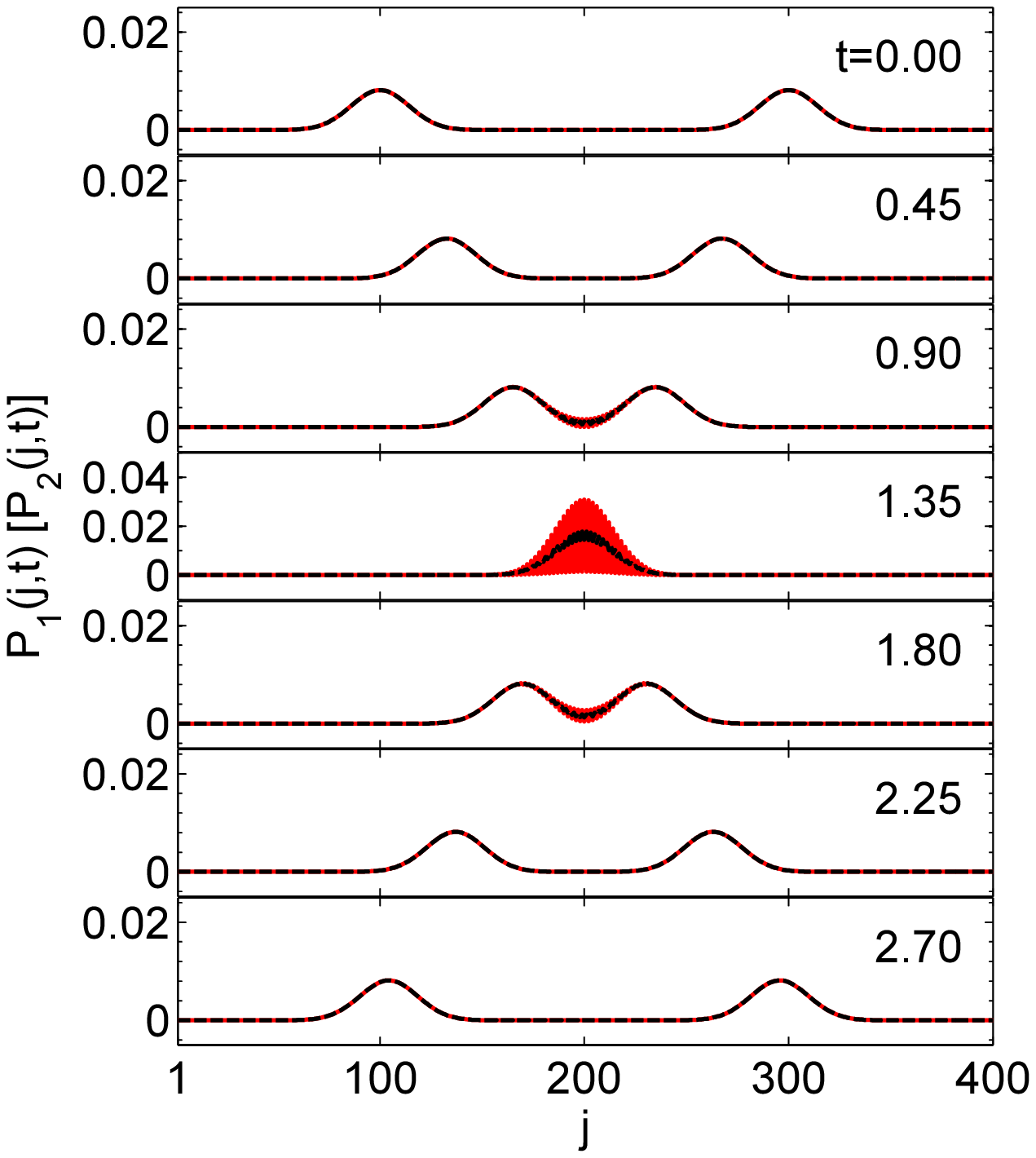} %
\includegraphics[ bb=30 185 403 601, width=6.5 cm, clip]{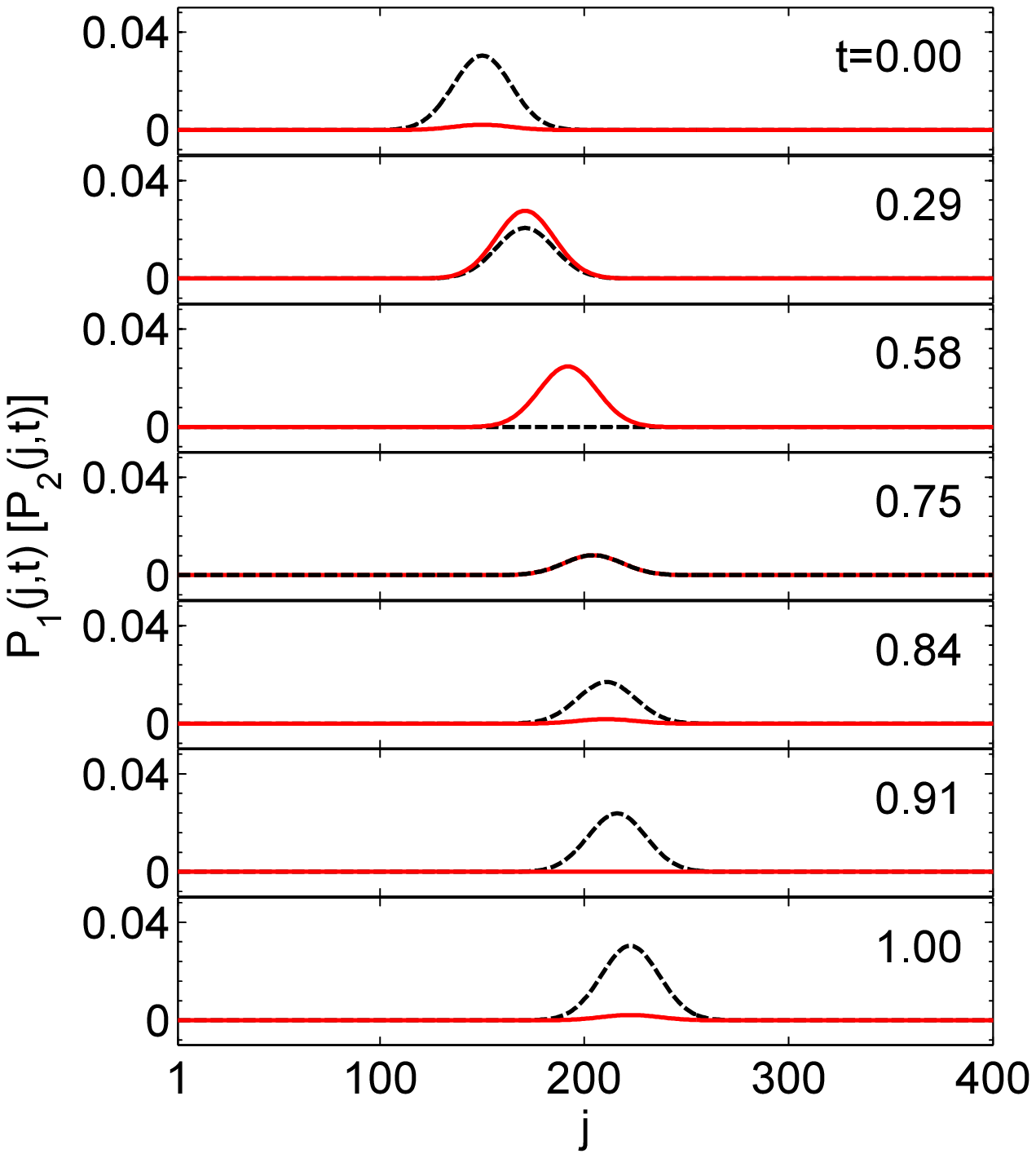}
\caption{(Color online) The Dirac probabilities $P_{1}\left( j,t\right) $
(black dashed line) and $P_{2}\left( j,t\right) $ (red solid line) of a
particle, initially located in the state $\left\vert \Psi \left( N_{A},N_{B},%
\phi _{A}, \phi _{B},0\right) \right\rangle $ for a system
with $N=400$, $\gamma =0.05,J=0.10, \kappa =1.00$, and $%
\rho =0.05$. We obtain $\theta =\pi /6$ and the time
$t$ is in units of $T_{\text{D}}\approx 36.276$ $\kappa ^{-1}$. We
plot the Eq. (\ref{P_l}) for two cases with (a) $\phi _{A}=-%
\phi _{B}=\pi /2$, $N_{A}=100$, $N_{B}=300$ and (b) $%
\phi _{A}=\phi _{B}=\pi /2$, $N_{A}=N_{B}=150$. The shapes
of all the curves are in agreement with the analysis in the text.}
\label{fig_evol}
\end{figure*}


Considering an arbitrary state involving both upper and lower bands

\begin{equation}
\left\vert \Phi \left( 0\right) \right\rangle =\sum_{k,\sigma =\pm
}f_{k,\sigma }\bar{a}_{k,\sigma }\left\vert \text{vac}\right\rangle ,
\end{equation}%
with $\sum_{k,\sigma =\pm }\left\vert f_{k,\sigma }\right\vert ^{2}=1$, we
have%
\begin{eqnarray}
\left\vert \left\vert \Phi \left( t\right) \right\rangle \right\vert
^{2}=\sum_{k,\sigma }\left\vert f_{k,\sigma }\right\vert ^{2}\left\langle
\text{vac}\right\vert [\bar{a}_{k,\sigma }^{\dag },\bar{a}_{k,\sigma }]_{\pm
}\left\vert \text{vac}\right\rangle &&  \label{Phi_t} \\
+\sum_{k,\sigma }f_{k,-\sigma }^{\ast }f_{k,\sigma }e^{-i2\sigma \Delta
t}\left\langle \text{vac}\right\vert [\bar{a}_{k,-\sigma }^{\dag },\bar{a}%
_{k,\sigma }]_{\pm }\left\vert \text{vac}\right\rangle &&  \notag \\
=\sec \theta +i\tan \theta \sum_{k,\sigma }\sigma f_{k,-\sigma }^{\ast
}f_{k,\sigma }e^{-i\sigma 2\pi (t/T_{\text{D}})}, &&  \notag
\end{eqnarray}%
where $T_{\text{D}}=\pi /\Delta $ denotes the period of the oscillation. The
first term gives the contribution from single band, while the second term
captures the influence of the non-Hermiticity. For vanishing $\theta $\ we
recover the unitary evolution in Hermitian system. Evidently, $\left\vert
\left\vert \Phi \left( t\right) \right\rangle \right\vert ^{2}=\sec \theta $
for a state with $f_{k,-\sigma }^{\ast }f_{k,\sigma }=0$, which involves
only a single mode. Note, however, that mathematically speaking the time
dependent terms can vanish even in the case of $f_{k,-\sigma }^{\ast
}f_{k,\sigma }\neq 0$, e.g. additional orthogonality of the wavepacket with
multiple eigenmodes. To demonstrate this, we study the evolution of initial
wavepackets of the form
\begin{eqnarray}
&&\left\vert \Psi \left( N_{A},N_{B},\phi _{A},\phi _{B},0\right)
\right\rangle =\frac{1}{\sqrt{\Omega }}  \label{GWP_k_space} \\
&&\times \sum_{k}\left[ e^{-\left( k-\phi _{A}\right) ^{2}/(2\rho
^{2})}e^{-i\left( k-\phi _{A}\right) N_{A}}\bar{a}_{k,+}\right.  \notag \\
&&+\left. e^{-\left( k-\phi _{B}\right) ^{2}/(2\rho ^{2})}e^{-i\left( k-\phi
_{B}\right) N_{B}}\bar{a}_{k,-}\right] \left\vert \text{vac}\right\rangle ,
\notag
\end{eqnarray}%
which is the superposition of wavepackets $A$ and $B$, where $\Omega
=2\sum_{k}e^{-\left( k-\phi _{A}\right) ^{2}/\rho ^{2}}$ $%
=2\sum_{k}e^{-\left( k-\phi _{B}\right) ^{2}/\rho ^{2}}$. The time evolution
of wavepacket is a powerful tool for understanding the dynamical property of
Hermitian quantum systems \cite{Littlejohn}. Recently, The propagation of
wavepacket in discrete systems has been utilized as flying qubit for quantum
state transfer \cite{Linden,ST,SYang73,LJin042341,ZXZ}. In the Hermitian
case an initially Gaussian state stays Gaussian as it propagates for a long
time, especially for the case of $\left\vert \phi _{A,B}\right\vert =\pi /2$
\cite{WKim}.

For a sufficient broad wavepacket ($\rho \ll 1$), we have $\Omega \approx
\rho N/\sqrt{\pi }$. Equation (\ref{GWP_k_space}) can also be expressed in
the coordinate space spanned by \{$a_{\alpha ,1}^{\dagger
}\left\vert \text{vac}\right\rangle $, $a_{\alpha ,2}^{\dagger }\left\vert
\text{vac}\right\rangle $\} as

\begin{eqnarray}
&&\left\vert \Psi \left( N_{A},N_{B},\phi _{A},\phi _{B},0\right)
\right\rangle \approx \sqrt{\frac{\rho }{4\sqrt{\pi }\cos \theta }}
\label{GWP_leg} \\
&&\times \sum_{\alpha =1}^{N}\left[ e^{-\rho ^{2}\left( j-N_{A}\right)
^{2}/2}e^{i\phi _{A}j}\right. \left( e^{i\theta /2}a_{\alpha ,1}^{\dagger
}-e^{-i\theta /2}a_{\alpha ,2}^{\dagger }\right)  \notag \\
&&+\left. e^{-\rho ^{2}\left( j-N_{B}\right) ^{2}/2}e^{i\phi _{B}j}\left(
e^{-i\theta /2}a_{\alpha ,1}^{\dagger }+e^{i\theta /2}a_{\alpha ,2}^{\dagger
}\right) \right] \left\vert \text{vac}\right\rangle ,  \notag
\end{eqnarray}%
which involves both eigenmodes ($\sigma =\pm $) and\ actually composed of
four wavepackets with centers at $N_{A}$th and $N_{B}$th sites of the
legs $1$ and $2$, and with the velocities $\phi _{A}$ and $\phi _{B}$,
respectively. To investigate the dynamics of the Dirac norm, substituting
\begin{subequations}
\begin{eqnarray}
f_{k,+} &=&\frac{1}{\sqrt{\Omega }}e^{-\left( k-\phi _{A}\right)^{2}/(2\rho ^{2})}e^{-i(k-\phi _{A})N_{A}},\\
f_{k,-} &=&\frac{1}{\sqrt{\Omega }}e^{-\left( k-\phi _{B}\right)^{2}/(2\rho ^{2})}e^{-i(k-\phi _{B})N_{B}},
\end{eqnarray}%
\end{subequations}
into Eq. (\ref{Phi_t}), we have

\begin{eqnarray}
&&\left\vert \left\vert \Psi \left( N_{A},N_{B},\phi _{A},\phi _{B},t\right)
\right\rangle \right\vert ^{2}=\sec \theta + \\
&&e^{-\left( \phi _{A}-\phi _{B}\right) ^{2}/(4\rho ^{2})}e^{-\rho
^{2}\left( N_{B}-N_{A}\right) ^{2}/4}\sin \left( 2\pi t/T_{\text{D}}-\varphi
_{AB}\right) \tan \theta .  \notag
\end{eqnarray}%
where $\varphi _{AB}=$ $\left( N_{A}+N_{B}\right) \left( \phi _{A}-\phi
_{B}\right) /2$. We note that if the two wavepackets of Eq. (\ref%
{GWP_k_space}) are well separate in $k$ or $\alpha $\ space initially
(wavepackets orthogonal in $k$\ or $\alpha $\ space), the weighted
exponential factor becomes zero, then the probability is always conserved in
the evolution even they meet each other in the coordinate space $\alpha $. This
indicates that for states having additional orthogonal modes, Hermitian like
behavior still holds even multiple eigenmodes are involved.

To show more detailed propagation behavior, we study the profile of $%
P_{\ell }\left( j,t\right) $\ ($\ell =1,2$), where%
\begin{equation}
P_{\ell }\left( j,t\right) =\left\vert \left\langle \text{vac}\right\vert
a_{j,\ell }\left\vert \Psi \left( N_{A},N_{B},\phi _{A},\phi _{B},t\right)
\right\rangle \right\vert ^{2},  \label{P_l}
\end{equation}%
It is a convenient way to investigate the dynamical properties from two
typical cases: (a) $\phi _{A}=-\phi _{B}=\pi /2$, $\left\vert
N_{A}-N_{B}\right\vert \gg 2\sqrt{\ln 2}/\rho $ and (b) $\phi _{A}=\phi
_{B}=\pi /2$, $N_{A}=N_{B}$. In case (a), the situation corresponds to two
counter-propagating wavepackets, with the evolved wave function

\begin{eqnarray}
&&\left\vert \Psi \left( N_{A},N_{B},\pi /2,-\pi /2,t\right) \right\rangle =%
\frac{1}{\sqrt{\Omega }} \\
&&\times \sum_{k}\left[ e^{-i\Delta t}e^{-\left( k-\pi /2\right) ^{2}/(2\rho
^{2})}e^{-i\left( k-\pi /2\right) \left( N_{A}+2\kappa t\right) }\bar{a}%
_{k,+}\right.  \notag \\
&&+e^{i\Delta t}\left. e^{-\left( k+\pi /2\right) ^{2}/(2\rho
^{2})}e^{-i\left( k+\pi /2\right) \left( N_{B}-2\kappa t\right) }\bar{a}%
_{k,-}\right] \left\vert \text{vac}\right\rangle  \notag \\
&=&\left\vert \Psi ^{\prime }\left( N_{A}+2\kappa t,N_{B}-2\kappa t,\pi
/2,-\pi /2,0\right) \right\rangle ,  \notag
\end{eqnarray}%
where the approximation of Taylor expansions for $\cos k$ around $\pm \pi /2$
are used for two wavepackets and $\left\vert \Psi ^{\prime }\right\rangle $
represents the superposition of two wavepackets as state $\left\vert \Psi
\right\rangle $ but with different overall phases. It shows that the evolved
state is still the independent nonspreading wavepackets. Similarly, the
evolved wave function for case (b) has the form

\begin{eqnarray}
&&\left\vert \Psi \left( N_{A},N_{A},\pi /2,\pi /2,t\right) \right\rangle =%
\frac{1}{\sqrt{\Omega }}\sum_{k}\left[ e^{-\left( k-\pi /2\right)
^{2}/(2\rho ^{2})}\right.  \notag \\
&&\left. e^{-i\left( k-\pi /2\right) \left( N_{A}+2\kappa t\right) }\left(
\bar{a}_{k,+}e^{-i\Delta t}+\bar{a}_{k,-}e^{i\Delta t}\right) \right]
\left\vert \text{vac}\right\rangle .
\end{eqnarray}%
It has more clear profile in the coordinate space $\ell $, i.e.%
\begin{eqnarray}
&&\left\vert \Psi \left( N_{A},N_{A},\pi /2,\pi /2,t\right) \right\rangle \\
&\approx &\sum_{\ell =1,2}g_{\ell }\left( t\right) \sum_{j=1}^{N}e^{-\rho
^{2}\left[ j-\left( N_{A}+2\kappa t\right) \right] ^{2}/2}e^{ij\pi
/2}a_{j,\ell }^{\dagger }\left\vert \text{vac}\right\rangle ,  \notag
\end{eqnarray}%
where%
\begin{equation}
g_{\ell }\left( t\right) =\sqrt{\frac{\rho }{\sqrt{\pi }\cos \theta }}\times
\left\{
\begin{array}{c}
\cos \left( \pi t/T_{\text{D}}-\theta /2\right) ,\ell =1 \\
i\sin \left( \pi t/T_{\text{D}}+\theta /2\right) ,\ell =2%
\end{array}%
\right. .
\end{equation}%
Obviously, it represents two breathing shape-invariant wavepackets
propagating along two legs of the ladder with the breathing period $T_{%
\text{D}}$. Furthermore, the Dirac norm $P_{\ell }^{s}=\sum_{j}P_{\ell
}\left( j,t\right) $ ($\ell =1,2$) and $P_{\text{T}}^{s}=P_{1}^{s}+P_{2}^{s}$
can be obtained as the form%
\begin{eqnarray}
P_{1}^{s} &=&\cos ^{2}\left( \pi t/T_{\text{D}}-\theta /2\right) /\cos
\theta , \\
P_{2}^{s} &=&\sin ^{2}\left( \pi t/T_{\text{D}}+\theta /2\right) /\cos
\theta , \\
P_{\text{T}}^{s} &=&\sec \theta +\tan \theta \sin \left( 2\pi t/T_{\text{D}%
}\right) .
\end{eqnarray}%
As mentioned in the introduction, the profile of the evolved wave function $%
P_{\ell }\left( j,t\right) $\ can be observed in experiment. In practice,
the quantum-optical analogy has been employed to visualize the dynamics in
the non-Hermitian system \cite{PRL103904,PRL030402,OL2632}. In this context,
the light intensity corresponds to $P_{\ell }\left( j,t\right) $\ (for a
review, see \cite{LonghiRev}) and the profile corresponds to the light
intensity distribution along its propagation direction.

It follows that a manifestation of the non-Hermitian nature of $H_{\text{Ladd%
}}$\ is represented by the relative phase $\theta $\ between the breathing
oscillations of the two legs, which also leads to the time-dependent Dirac
probability.
\begin{figure}[tbp]
\includegraphics[ bb=34 178 548 590, width=6.8 cm, clip]{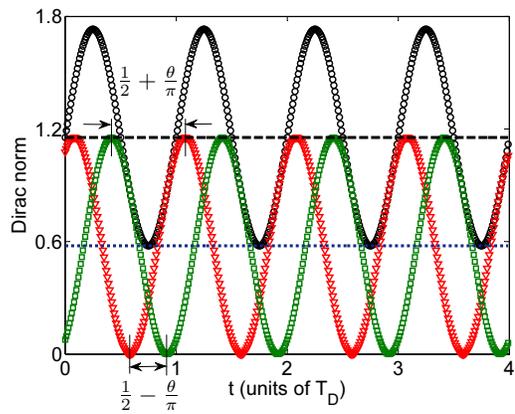}
\caption{(Color online) The Dirac norms $P_{1}^{s}\left( t\right) $, $P_{2}^{s}\left( t\right) $ (blue dotted line) and $P_{\text{T}}^{s}\left(
t\right) $ (black dashed line) for the case of $\phi _{A}=-\phi _{B}=\pi /2$ [as in Fig. \ref{fig_evol}(a)]. The Dirac
norms $P_{1}^{s}\left( t\right) $ (red triangle), $P_{2}^{s}\left( t\right) $
(green square) and $P_{\text{T}}^{s}\left( t\right) $ (black circle) for the
case of $\phi _{A}=\phi _{B}=\pi /2$ [as in Fig.
\ref{fig_evol}(b)]. All the parameters are the same as in Fig.
\ref{fig_evol}. The phase difference $\theta=\pi/6$
and also the quasi-canonical commutation relations $\sec\theta\approx1.155$ are indicated . The shapes of all the curves are in agreement
with the analysis in the text.} \label{fig_Pro}
\end{figure}
The profiles of the evolved wave functions and the Dirac norms are plotted
in Figs. \ref{fig_evol} and \ref{fig_Pro}. We can see that in case (a) the
evolved wavepackets propagate independently and the Dirac norms are
preserving. It indicates that although the Hamiltonian is non-Hermitian, due
to the quasi-canonical commutation relations which is a direct consequence
of the Hermitian connection structure between clusters, it acts as a
Hermitian ladder for some initial state. In contrast, the dynamics of
case (b) differs drastically from the Hermitian case and the Dirac norm is
no long preserved. Further, the phase difference between the breathing
oscillations on the two legs can also be observed in case (b).

\section{Summary and discussions}
\label{sec_conclusion} In summary, we show in this paper within the context
of a class of non-Hermitian lattice systems, which consist of a set of
isomorphic pseudo-Hermitian clusters combined in a Hermitian manner, that
Hermitian like dynamics could be observed in such non-Hermitian systems,
including the property that the time evolution is Dirac probability
preserving. As an application, we investigate a concrete network, a $%
\mathcal{PT}$-symmetric ladder, composed of many pseudo-Hermitian dimers.
It is shown that it acts as a Hermitian system in the following sense:
besides the reality of the spectrum and probability preserving, the
propagation of certain wavepackets exhibit the same behavior as that in a
Hermitian ladder. Our finding indicates that the reality of the spectrum as
well as the Dirac probability preserving dynamics can occur in a system that
violating the axiom of Hermiticity. This will pave the way for the
development of descriptions of quantum system and provide a topic of
considerable interest in a wide range of subjects.

\acknowledgments We acknowledge the support of the CNSF (Grant No. 10874091)
and National Basic Research Program (973 Program) of China under Grant No.
2012CB921900.

\end{document}